\documentclass{emulateapj}

\usepackage{amsmath}

\def\msun{{\rm M_{\odot}}}

\def\be{\begin{equation}}
\def\ee{\end{equation}}

\def\le{{L_{\rm Edd}}}
\def\msun{{\rm M_{\odot}}}

\def\mo{{\dot M_{\rm out}}}
\def\me{{\dot M_{\rm Edd}}}

\begin{document}

\title{THE AGN--STARBURST CONNECTION, GALACTIC SUPERWINDS, AND
 $M_{\rm BH} - \sigma$}

\author{ Andrew~King\altaffilmark{1}}

\altaffiltext{1} {Department of Physics and Astronomy, University of
Leicester, Leicester LE1 7RH, U.K.; ark@astro.le.ac.uk}

\begin{abstract}
Recent observations of young galaxies at redshifts $z \sim 3$ have revealed
simultaneous AGN and starburst activity, as well as galaxy--wide superwinds.
I show that there is probably a close connection between these phenomena by
extending an earlier treatment of the $M_{\rm BH} - \sigma$ relation (King,
2003). As the black hole grows, an outflow drives a shell into the surrounding
gas. This stalls after a dynamical time at a size determined by the hole's
current mass and thereafter grows on the Salpeter timescale.  The gas trapped
inside this bubble cools and forms stars and is recycled as accretion and
outflow. The consequent high metallicity agrees with that commonly observed in
AGN accretion. Once the hole reaches a critical mass this region attains a
size such that the gas can no longer cool efficiently. The resulting
energy--driven flow expels the remaining gas as a superwind, fixing both the
$M_{\rm BH} - \sigma$ relation and the total stellar bulge mass at values in
good agreement with observation. Black hole growth thus produces starbursts
and ultimately a superwind.

\end{abstract}


\section{Introduction}

Most astronomers now believe that the centre of every galaxy contains
a supermassive black hole. There is a close observational correlation
(Ferrarese \& Merritt, 2000; Gebhardt et al., 2000; Tremaine et al.,
2002) between the mass $M$ of this hole and the velocity dispersion
$\sigma$ of the host bulge. This strongly suggests a connection
between the formation of the black hole and of the galaxy
itself. 

There have been many attempts to explain this correlation
theoretically. Some of these appeal to the ambient conditions in the
host galaxy (Adams, Graff \& Richstone, 2001), accretion of
collisional dark matter (Ostriker, 2000), or star captures by the
central accretion disc (Miralda--Escud\'e \& Kollmeier, 2005). A very
large class of models (Silk and Rees 1998; Haehnelt et al., 1998;
Blandford, 1999; Fabian 1999; Wyithe \& Loeb, 2003; King, 2003; Murray
et al., 2005; Sazonov et al., 2005; Robertson et al. 2005; Di Matteo et
al., 2005; Begelman \& Nath, 2005) use the idea of feedback by a wind
or outflow driven in some way by the accretion required to grow the
central black hole. In this picture, the black hole eventually reaches
a mass $M_{\rm BH}$ such that further accretion (and thus growth of
$M_{\rm BH}$) is prevented because the outflow sweeps away the ambient
gas.

This is a very natural idea, given that we know that most of the mass
of the nuclear black holes is assembled by luminous accretion (Soltan
1982; Yu \& Tremaine, 2002). Moreover, the rate at which mass tries to
flow in towards the central black hole in a galaxy is set by
conditions far away, for example by interactions or mergers with other
galaxies. These rates must be at least comparable with the instantaneous
Eddington rates in order to produce the known black hole masses in the
available time (particularly at higher redshifts). Since these
external conditions cannot know what the current mass of the central
black hole is, it seems very likely that super--Eddington conditions
prevail for most of the time that the central black hole mass grows,
and indeed this is what studies of the cosmological history of the
process suggest (cf Miller et al., 2005).

It is therefore reasonable to consider the effect on the host galaxy
of such super--Eddington accretion. Simple theory (King \& Pounds,
2003) motivated by X--ray observations of outflows from bright quasars
(e.g. Pounds et al., 2003a, b; Reeves et al., 2003)
suggests that the outflow momentum flux is comparable to that of the
Eddington--limited radiation field, i.e.
\begin{equation}
\mo v \simeq {\le\over c},
\label{mom}
\end{equation}
where $\mo$ is the mass outflow rate and $\le$ the Eddington
luminosity, while the mechanical energy flux is
\begin{equation}
{1\over 2}\mo v^2 \simeq {\le^2\over 2\mo c^2} 
\label{en}
\end{equation}

The wind from the central black hole sweeps up the surrounding gas into a
shell. The theory of stellar wind bubbles (e.g. Lamers \& Casinelli 1999)
shows that this shell is bounded by an inner shock where the wind velocity is
thermalized, and an outer shock where the surrounding gas is heated and
compressed by the wind. These two regions are separated by a contact
discontinuity. The dynamics of the feedback process are then determined by
whether the swept--up gas can cool (momentum--driven flow) or not
(energy--driven flow) on a timescale shorter than the flow time. In fact
Compton cooling is likely to be effective (King, 2003: hereafter K03) until a
very large (galaxy--scale) mass has been swept up (see eq. \ref{mr}
below). Then eq. (\ref{mom}) states that the effect of super--Eddington
accretion is to inject the Eddington momentum into the surroundings. In a
simple picture, no further accretion on to the black hole is possible once tha
shell of swept--up matter has acquired the escale velocity $\sim \sigma$. The
resulting theory then has no free parameter: remarkably, it leads to an $M -
\sigma$ relation $M_{\rm BH} \propto \sigma^4$ very close to the observed one
($M_8 \simeq \sigma_{200}^4$, with $M_8 = M_{\rm BH}/\msun,\ \sigma_{200} =
\sigma/200\ {\rm km\ s^{-1}}$) in both slope and normalization (K03).

Recent observations now make it possible to check the consequences of
this simple model further. In particular there is good reason to
believe (Alexander et al, 2005) that the growth of the black hole mass
(revealed by AGN activity) occurs simultaneously with the formation of
stars in the spheroid. Moreover there is now suggestive evidence of
very large--scale outflows from young galaxies at redshifts $z \sim 3$
(Wilman et al., 2005; see also Binette et al., 2000). A fuller
treatment of the problem is therefore appropriate. I begin by
considering the initial momentum--driven outflow in more detail.

\section{Black Hole Growth}

We model a protogalaxy as an isothermal sphere of
dark matter. If the gas fraction is $f_g = \Omega_{\rm
baryon}/\Omega_{\rm matter}\simeq 0.16$ (Spergel et al., 2003) its
density is
\begin{equation}
\rho = {f_g\sigma^2\over 2\pi Gr^2}
\label{rho}
\end{equation}
where $\sigma$ is assumed constant. The gas mass inside radius $R$ is
\begin{equation}
M(R) = 4\pi\int_0^R\rho r^2 {\rm d}r = {2f_g\sigma^2R\over G}
\label{m}
\end{equation}
so this is the mass of the shell of swept--up matter at radius $R$.
Inside a certain radius $R_c$ (see below) the shocked wind can always
cool so we have an outflow driven by the momentum rate (\ref{mom})
\begin{equation}
\mo v \simeq {\le\over c},
\end{equation}
where $\mo$ is the mass outflow rate and $\le$ the Eddington
luminosity. 

K03 simplified the problem by considering only the driving by this force, and
asking under what conditions the resulting outflow velocity reaches the escape
value $\sim \sigma$. To follow the process in detail we must include the
resistance by gravity. The equation of motion is
\begin{eqnarray}
{{\rm d}\over {\rm d}t}[M(R)\dot R] + {GM(R)[M + M(R)]\over R^2} = 
4\pi R^2\rho v^2 \hspace{10pt}\nonumber\\
= \mo v = {\le\over c}
\label{motion}
\end{eqnarray}
where we have used the mass conservation equation $4\pi R^2\rho v =
\mo$ for the quasar wind velocity $v$, and then (\ref{mom}) to
simplify the rhs. Using (\ref{m}) we
can simplify the gravity term, so
\begin{equation} 
{{\rm d}\over {\rm d}t}\biggl[M(R)\dot R\biggr] +
2f_g{\sigma^2\over R}\biggl[M + {2\sigma^2R\over
G}\biggr] = {\le\over c}.
\label{motion2}
\end{equation}
Then using (\ref{m}) on the lhs we get finally
\begin{equation}
{{\rm d}\over {\rm d}t}(R\dot R) + {GM\over R} =
-2\sigma^2\biggl[1 - {M\over M_{\sigma}}\biggr]
\label{motion3}
\end{equation}
where
\begin{equation}
M_{\sigma} = {f_g\kappa\over \pi G^2}\sigma^4.
\label{msig}
\end{equation}
It is easy to show from (\ref{motion3}) that for $M < M_{\sigma}$ we
have $\ddot R < 0$, i.e. the shell always decelerates.

Now let us consider the growth of the black hole mass $M$. As argued above,
this is likely to be super--Eddington at small $M$ and set by conditions far
from the hole. Thus any feedback ultimately limiting the growth of $M$ can
only occur when the bubble radius $R$ is similarly large. This requires the
wind outflow to be launched with at least the local escape velocity near the
black hole. X--ray observations of outflows indeed suggest this (see the
discussion in King \& Pounds, 2003). For $R \gg GM/\sigma^2$ we
may neglect the black--hole gravity term $GM/R$ in (\ref{motion3}). Then
integrating twice gives
\begin{equation}
R^2 = R_0^2 + 2R_0\dot R_0t - 2\sigma^2\biggl[1 - {M\over
M_{\sigma}}\biggr]
\label{r}
\end{equation}
where $\dot R = \dot R_0$ at $R = R_0$ with $R_0$ some large radius. One could
in principle connect $R_0, \dot R_0$ to the initial conditions near the black
hole by numerical integration of the full equation (\ref{motion3}), but this
is unnecessary for our purposes. From (\ref{r}) we see that for $M <
M_{\sigma}$ the shell reaches a maximum radius $R_{\rm max}$ given by
\begin{equation}
{R_{\rm max}^2\over R_0^2} = 1 + {\dot R_0^2\over 2\sigma^2[1 -
M/M_{\sigma}]}
\label{rm}
\end{equation}
after a time
\begin{equation}
t_{\rm max} = {R_0\dot R_0\over 2\sigma^2[1 - M/M_{\sigma}]}
\label{tm}
\end{equation}
before stalling. Thus the bubble reaches a maximum size $R_{\rm max}$
in a dynamical time. For consistency we require $GM/R_{\rm max} \ll
\sigma^2$, or that the swept--up mass 
\begin{equation}
M_{\rm max} = M(R_{\rm max}) = {2f_g\sigma^2\over G}R_{\rm max} \gg
2f_gM
\end{equation}
should be rather larger than the black hole mass.

Once the bubble reaches the mass $M_{\rm max}(M)$, the ram pressure of the
outflow cannot sweep up any additional mass. Thus in particular the shocked
wind gas cannot add to the shell, but must fall back and orbit as it
cools. Some of this gas forms stars and begins to grow the spheroidal stellar
component. Depending on how angular momentum is redistributed, some of the
shocked wind gas is instead potentially available to continue refuelling the
black hole, very possibly at a super--Eddington rate. Thus if $\mo \gg \me$,
gas may be repeatedly recycled as outflow while the black hole mass $M$ and
the bubble radius $R$ grow on the Salpeter timescale $t_s = M/\me$, increasing
$M_{\max}(M)$ on the same timescale. The most massive stars have lifetimes
less than $t_s$, so much of their gas, now enriched in metallicity, is
returned to the ISM and is potentially available for feeding the hole once
more.

The continuous compression and recycling occurring during the black hole
growth must promote efficient star formation. Hence this picture indeed
explains why AGN activity and starbursts are simultaneous, as in observations
of SCUBA galaxies by Alexander et al. (2005). Moreover it accounts for the
high metallicity commonly observed in gas accreting in AGN. If the supply of
new gas for accretion (e.g. produced by merger events) is slow, star formation
will deplete the gas supply to the point where accretion gradually shuts
off. The swept--up gas (by now mostly turned into stars) must fall back, thus
forming a spheroidal stellar bulge around the black hole.  Such a galaxy would
have $M_{\rm BH}$ lying below the $M - \sigma$ relation, only growing towards
it again as a new merger event supplies more mass.

\section{The $M - \sigma$ relation}

Given an adequate mass supply (e.g. by mergers), $M$ grows on the Salpeter
timescale as explained above. However the shell radius $R_{\rm max}$ gets very
large (whatever the values of $R_0, \dot R_0$) as $M$ approaches the value
where the denominator in (\ref{rm}) vanishes, i.e.
\begin{equation}
M_{\rm BH} = {f_g\kappa\over \pi G^2}\sigma^4 
\label{mcrit}
\end{equation}
where $\kappa$ is the electron scattering opacity. In particular, at
such masses $R_{\rm max}$ reaches the radius $R_c$ at which Compton
losses no longer cool the shocked wind efficiently (see K03). At this
point the extra pressure accelerates the shell so that it escapes the
galaxy entirely (see below), thus finally shutting off accretion.

Thus the black hole mass cannot grow beyond that given by (\ref{mcrit}), and
this is the $M - \sigma$ relation. The proportionality constant is twice that
of the simple treatment in K03, but still well within observational scatter.

\section{The Bulge Mass}

If the shocked gas no longer efficiently cools then the driving force
on the shell is now the total gas pressure $P$, which is larger than
the ram pressure $\rho v^2$ appearing in (\ref{motion}). The equation
of motion thus becomes
\begin{equation}
{{\rm d}\over {\rm d}t}\biggl[M(R)\dot R\biggr] + {GM(R)M_(R)\over R^2} = 
4\pi R^2P
\label{motion2}
\end{equation}
We now have to add the energy equation, allowing for the rate that
energy is fed into the shocked gas, minus the rate of work on the
ambient gas and against gravity:
\begin{equation}
{{\rm d}\over {\rm d}t}\biggl[{4\pi R^3\over 3}.{3\over 2}P\biggr] = 
{\eta\over 2}\le - P{{\rm d}\over {\rm d}t}\biggl({4\pi\over
  3}R^3\biggr)
- 4f_g{\sigma^4\over G}
\label{energy}
\end{equation}
where I have used (\ref{en}) to specify energy input from the outflow,
$\eta \simeq 0.1$ is the accretion efficiency, and I have used
(\ref{m}) to simplify the gravity term $GM(R)M(R)/R^2$.

Now using (\ref{motion2}) to eliminate $P$ from (\ref{energy}), and
replacing the gravity terms as before, we get
\begin{eqnarray}
{\eta\over 2}\le = \dot R{{\rm d}\over {\rm d}t}\biggl[M(R)\dot R\biggr]
+ 8f_g{\sigma^4\over G}\dot R + \hspace{35pt}\nonumber\\
{{\rm d}\over {\rm d}t}\biggl\{{R\over 2}\dot R{{\rm d}\over {\rm d}t}
\biggl[M(R)\dot R\biggr] + 2f_g{\sigma^4\over G}R\biggr\}
\end{eqnarray}
leading to
\begin{eqnarray}
{\eta\over 2}\le = 
{2f_g\sigma^2\over G}\biggl\{ {1\over 2}R^2\dddot R + 3R\dot R\ddot R
+ \hspace{45pt}\nonumber\\
{3\over 2}\dot R^3\biggr\} + 10f_g{\sigma^4\over G}\dot R
\label{motion4}
\end{eqnarray}
Using $M= M_{\rm BH}$ from ({\ref{mcrit}) in $\le$, this has a
solution $R = v_e t$ with
\begin{equation}
 \eta c = 3{v_e^3\over \sigma^2} + 10v_e
\end{equation}
The assumption $v_e \ll \sigma$ leads to a contradiction ($v_e \simeq
0.01c \gg \sigma$), hence this equation has the approximate solution
\begin{equation}
v_e \simeq \biggl[{2\eta\sigma^2c\over 3}\biggr]^{1/3} \simeq
875\sigma_{200}^{2/3}~{\rm km\ s}^{-1}
\label{ve}
\end{equation}

Hence the remaining gas outside $R_c$ is driven off completely. The
total gas mass turned into bulge stars is thus
\begin{equation}
M_b = M(R_c) = 1.9\times 10^{11}\sigma_{200}^3M_8^{1/2}
\biggl({v\over c}\biggr)^2b^{-1}f_g^{3/2}~\msun
\label{mr}
\end{equation}
(from K03). As shown in K03, this stellar mass reproduces the
approximate relation $M_{\rm BH} \simeq 10^{-3}M_b$ (Magorrian et al.,
1998) if $f_g, v, b$ all differ little between galaxies. The $M_b -
M_{\rm BH}$ relation is known to be rather weaker than the $M_{\rm BH} -
\sigma$ relation, suggesting that there is indeed some spread
here. Equation (\ref{mr}) would give $M_b \sim \sigma^5$ if $f_g,
v, b$ were strictly the same for all galaxies, formally
slightly inconsistent with the Faber--Jackson relation. This again
suggest these quantities vary somewhat between galaxies.

\section{Superwinds}

We have seen that SMBH growth is likely to drive a bubble into the surrounding
gas and initiate vigorous star formation. We noted above that the lifetime of
massive stars is less than the Salpeter time, so the supernovae produced by
these stars must drive vigorous mass loss from within the bubble. Moreover
equation (\ref{ve}) shows that once the black hole mass reaches the value
given by the $M -\sigma$ relation (\ref{mcrit}) its Eddington luminosity can
in principle blow away all the mass beyond the cooling radius. It is tempting
to see this as the origin of some observed galaxy--wide superwinds.  

In particular it is worth noting that superwind driving by outflows from SMBH
as discussed here is more energy--efficient than the conventional picture of
supernova driving. Thus in the example discussed by Wilman et al. (2005) the
shell has mass $\sim 10^{11}\msun$ and velocity $\sim 250~{\rm km~s}^{-1}$ and
hence kinetic energy $\sim 6.3\times 10^{58}$~erg. Supernovae provide $\sim
10^{60}$~erg over $10^8$~yr. However, the superwind is expected to cool from
the shock temperature $\sim 10^7$~K to a few $10^4$~K, in photoionization
equilibrium with the metagalactic ultraviolet radiation field, as it sweeps up
the intergalactic medium around the original host galaxy. It thus becomes
momentum--driven once again. This means that the conversion of supernova input
energy to kinetic energy is likely to be much lower than the required
6.3\%. By contrast, the Eddington momentum flux $\le/c$ for a $\sim 10^8\msun$
black hole operating for $\sim 10^8$~yr would provide a total momentum a few
times $10^{51}$~g~cm~s$^{-1}$, very close to the estimated value.

This paper shows that SMBH growth simultaneously produces starbursts and
strong outflows. Both are agents of mass loss, but the second is much more
efficient than the first. Thus it may be worth looking for signs of obscured
AGN activity where superwinds are seen and the accompanying starburst does not
seem sufficient to drive them

\section{Discussion}

I have studied the feedback probably responsible for the $M_{\rm BH} - \sigma$
relation in greater detail than in K03. In particular I allow for the effects
of gravity in slowing and ultimately stalling the swept--up shell of matter
around the growing black hole. The gas trapped within this stalled bubble is
efficiently turned into stars and can be recycled for accretion and outflow,
enhancing its metallicity as it does so. The bubble radius grows initially on
a dynamical time until it stalls, and then on the Salpeter timescale of the
central black hole. Once the bubble reaches the radius where it can no longer
efficiently cool, it carries away the gas further out in a fast,
energy--driven flow.

This simple picture accounts for a number of things. The stellar bulge
mass is the gas mass within the cooling radius, and agrees with the
Magorrian relation. The $M - \sigma$ relation follows because the
black hole cannot be efficiently fuelled once it is established. The
simultaneous growth of the stellar and black--hole masses agrees with
the observed AGN--starburst connection. The gas accreting in AGN must
have high metallicity because it is trapped in the stalled bubble and
repeatedly recycled through stars. Finally, the energy--driven outflow
ending black--hole growth suggests an explanation for galaxy--wide
superwinds where high efficiency is indicated.

Clearly the approach taken here is extremely simplified. In
particular, a fuller picture requires a better description of how
ambient gas loses enough angular momentum to accrete on to the central
black hole.

\acknowledgements

I thank David Alexander, Walter Dehnen and the referee for illuminating
discussions.  I gratefully acknowledge a Royal Society Wolfson
Research Merit Award.

\end{document}